# Spin-correlation Driven Ferroelectric Quantum Criticality in a Perovskite Quantum Spin-liquid System, $Ba_3CuSb_2O_9$

*S. Ghosh, G. Roy, E. Kushwaha, M. Kumar, and T. Basu**

*Rajiv Gandhi Institute of Petroleum Technology, Jais, Amethi 229304, India*

**Corresponding Author: tathamay.basu@rgipt.ac.in*

Here we have experimentally demonstrated spin-correlation-driven ferroelectric quantum criticality in a prototype quantum spin-liquid system, $Ba_3CuSb_2O_9$, a quantum phenomenon rarely observed. The dielectric constant follows a clear $T^2$ scaling, showing that the material behaves as a quantum paraelectric without developing ferroelectric order. Magnetically, the system avoids long-range order down to 1.8 K and instead displays a $T^{3/2}$ dependence in its inverse susceptibility, a hallmark of antiferromagnetic quantum critical fluctuations. Together with known spin-orbital-lattice entanglement in this compound, these signatures point to a strong interplay between spin dynamics and the polar lattice. Our pioneering work places this perovskite spin-liquid family at the forefront of this domain and suggest the flexibility of this family in a suitable environment by tuning chemical/ external pressure.

One of the most captivating ideas in modern condensed matter physics is that new phases of matter can arise not from order itself, but from the fluctuations that destabilize order [1,2]. This is the essence of a quantum critical point (QCP), a zero-temperature boundary where competing phases meet, tuned not by heat, but by non-thermal parameters such as pressure, doping, or external field [3-7]. Although theoretically the QCP exists strictly at absolute zero, practically its influence reaches far upward in temperature, creating a broad quantum critical regime, where fluctuations dominate the material's behaviour [1].

Magnetic quantum criticality has been extensively studied theoretically and experimentally over the last few decades, where ferromagnetic (FM) or antiferromagnetic (AFM) phase transitions are tuned towards absolute zero by parameters like external pressure, magnetic field, or chemical substitution. At such low temperatures, thermal fluctuations give way to quantum fluctuations that strongly influence the material's behavior [8,9]. Quantum critical phenomena have been reported in several correlated materials, each leading to distinct physical outcomes. For instance, $Sr_3Ru_2O_7$ exhibits a field-tuned metamagnetic quantum critical point [10], while $YbRh_2Si_2$ shows AFM quantum criticality associated with Kondo breakdown [11]. In $CeCu_{6-x}Au_x$, local-moment magnetic fluctuations dominate the quantum critical behavior [12], whereas $BaFe_2(As_{1-x}P_x)_2$ hosts a magnetic quantum critical point that is closely linked to unconventional superconductivity [13]. From the Hertz-Millis theoretical framework, the temperature dependence of the inverse magnetic susceptibility near a quantum critical point can be written as $\chi^{-1} \propto T^{(d+z-2)/z}$, where d is the spatial dimension of the system and z is the dynamical critical exponent. For a three-dimensional FM, the spatial dimension is $d = 3$. Because the spin fluctuations are strongly damped by the conduction electrons, the dispersion relation follows $\omega \sim q^3$, which gives z = 3. Substituting these values into the scaling relation leads to $\chi^{-1} \propto T^{4/3}$. In contrast, for a three-dimensional AFM, the spatial dimension is also d = 3, but the spin-fluctuation dispersion follows $\omega \sim q^2$, giving z = 2. Using these values in the same relation results in $\chi^{-1} \propto T^{3/2}$. Thus, within the Hertz-Millis theory, the inverse susceptibility shows a $T^{4/3}$ dependence for three-dimensional itinerant ferromagnets and a $T^{3/2}$ dependence for three-dimensional antiferromagnets near the quantum critical region [14,15].

For itinerant ferromagnets, the simple Hertz-Millis approach is incomplete because fermionic excitations and magnetic fluctuations remain strongly coupled, leading to first-order transitions or new phases instead of a clean quantum critical point. These scaling laws reflect the non-classical nature of spin dynamics in the quantum critical region. Studying such behavior helps in understanding how magnetism evolves under extreme quantum fluctuations in strongly correlated electron systems. On the other hand, ferroelectric quantum criticality (FE QC) has recently emerged as a central theme in understanding how quantum fluctuations influence the ground states of insulating oxides and related compounds [16]. Unlike quantum spin systems, where itinerant electrons complicate the picture, ferroelectrics offer a clear separation between electronic and phononic degrees of freedom and the lower complexity of many-body interactions, where criticality arises from soft optical phonons and polarization fluctuations of electric dipoles [17]. For three-dimensional FE, using Hertz-Millis theoretical framework, we can get d =3, and from the transverse phonon dispersion relation, it yields $\omega \sim q$, so z = 1, and using these values, we can get inverse dielectric susceptibility, $1/\varepsilon' \propto T^2$. Classical ferroelectrics such as $SrTiO_3$ and $KTaO_3$ display a characteristic $1/T^2$ divergence of the dielectric susceptibility near their QCP, consistent with theoretical models incorporating long-range dipolar forces and phonon coupling [18-20]. More complex cases, such as $BaFe_{12}O_{19}$, exhibit uniaxial ferroelectric quantum criticality with a $1/T^3$ scaling of susceptibility, reflecting a dimensional enhancement from long-range dipole interactions [21]. Chemical substitution and chemical pressure provide additional tuning in $Ba_{(1-x)}Ca_xFe_{12}O_{19}$, Ca substitution drives the system away from the QCP, stabilizing glassy or liquid-like dipolar states, with critical scaling of transition temperatures following quantum rather than classical exponents [22]. Beyond perovskites and hexaferrites, new families such as relaxor ferroelectrics (e.g., $K_3Li_2Ta_5O_{15}$ and $Pb_2Nb_2O_7$) demonstrate frequency-tuned relaxor-type QCPs, where dielectric freezing temperatures extrapolate below absolute zero, pointing to disordered ground states without long-range order [23]. Together, these studies highlight that ferroelectric QCPs can host unconventional dielectric responses, frustrated dipole liquids, expanding the paradigm of quantum criticality well beyond its magnetic analogues.

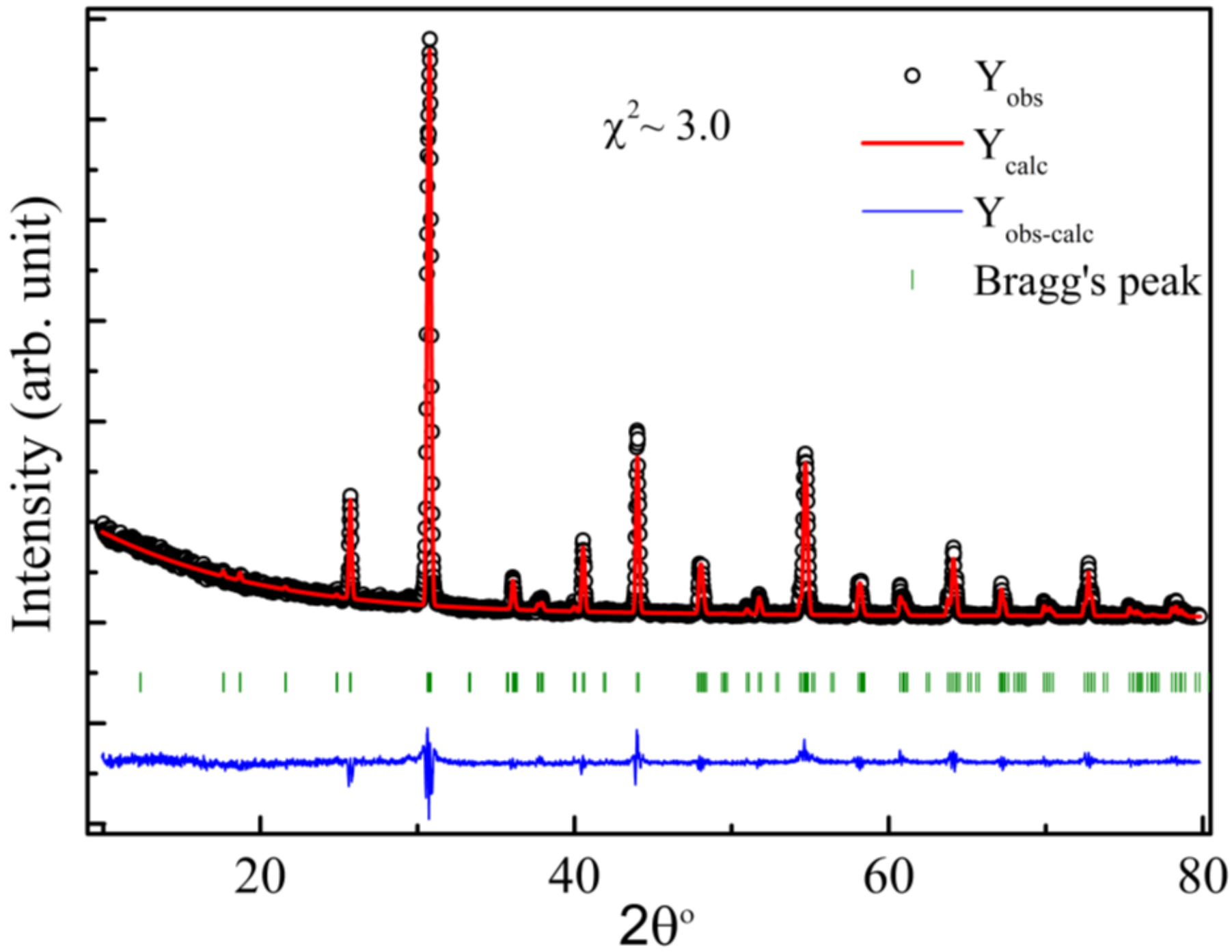

Figure 1 Rietveld refinement of the XRD pattern of BCSO at room temperature with space group $P6_3/mmc$.

In the recent developments of multiferroic systems where spins and dipoles are coupled to each other and exhibit fascinating phenomena such as spin-driven ferroelectricity in an ordered system [24,25]. A similar idea also appears that the coupling of electric and magnetic fluctuations might offer new pathways for emergent critical behaviour. Though briefer in appearance, these cases reflect the same unifying principle, when fluctuations govern the physics, entirely new states of matter can emerge. More recently, theorists have pushed the concept a step further by introducing the idea of multiferroic quantum criticality (MFQC) [14]. In this framework, electric and magnetic quantum fluctuations are not independent but entangled, allowing one order parameter to drive or amplify the other [26,27]. Dynamical multiferroicity, for example, predicts that oscillating electric dipoles can generate magnetic responses, even in systems with no static magnetism [27]. This raises a fascinating possibility, if both spin and ferroelectric degrees of freedom remain critical, then their interplay could stabilize new kinds of fluctuation-driven states, neither purely magnetic nor purely electric in nature, but inherently coupled to each other. Yet despite its rich science, experimental realizations of MFQC remain rare and limited to only one system, (Eu, Ba, Sr)$TiO_3$ [28], precisely because most materials find a way to order, either magnetically, orbitally, or structurally, before quantum fluctuations can fully take over. Here, we address this issue by investigating a prototype spin-liquid system experimentally.

The triangular-spin systems, $A_3MM'_2O_9$ (A = Ba, Sr, Ca; M = Ni, Co, Cu; M′ = Nb, Sb), exhibit several fascinating features arising from their triangular-lattice geometry and strong electronic correlations. [29-32] The presence of magnetic $M^{2+}$ ions on geometrically frustrated triangular planes leads to a delicate balance between exchange interactions and quantum fluctuations, giving rise to exotic magnetic ground states such as quantum spin-liquid, multiferroicity, etc [32-34]. Structural distortions induced by changing the A-site cation (Ba → Sr → Ca) or varying M′ (Nb/Sb) further tune the degree of frustration and anisotropy, allowing smooth crossover between magnetically ordered and disordered phases. For example, strong geometrical frustration and dominance of quantum fluctuation have been reported for compounds like, $Ba_3NiSb_2O_9$, $Ca_3NiNb_2O_9$, $Sr_3CuNb_2O_9$, $Sr_3CuSb_2O_9$ [32,35-38]. Whereas, in the same family, $Ba_3NiNb_2O_9$ demonstrates how quantum fluctuations in a triangular spin-1 lattice can stabilize exotic magnetic phases (including the up-up-down state), while simultaneously giving rise to multiferroicity (spin-driven ferroelectricity) across multiple phases [34]. Hence, this type of system could serve as a good candidate to investigate the quantum critical phenomena near the QCP regime, where spins and dipoles are coupled to each other. In $Ba_3NiNb_2O_9$, the spin-driven ferroelectricity is observed at the onset of long-range magnetic ordering at 4.9 K, even in the presence of quantum fluctuations [34]. Because of this long-range ordering, the phenomenon can't be considered purely quantum critical. In contrast, in $Ba_3CuSb_2O_9$ (BCSO), belonging to the same family, there is no long-range magnetic or ferroelectric order present. Our results suggest that in this compound the spin-driven ferroelectric behavior arises mainly from quantum fluctuations, which is consistent with a ferroelectric quantum critical scenario. In this manuscript, we investigated the hexagonal perovskite $Ba_3CuSb_2O_9$, which stands out as a unique candidate. In contrast to most $Cu^{2+}$ oxides, which show cooperative Jahn-Teller (JT) distortions and develop static orbital order, BCSO resists this tendency. Instead, it maintains its hexagonal symmetry to the lowest measured temperatures [39,40]. This concept emerges from the balance of spin, orbital, and lattice degrees of freedom, which remain dynamically entangled rather than freezing out. As a result, BCSO does not exhibit conventional magnetic or orbital order. Instead, it hosts a spin-orbital entangled liquid, a state where S = 1/2 spins on a disordered honeycomb lattice continue to fluctuate, coupled to dynamic JT distortions and orbital motion [40]. Neutron scattering, muon spin relaxation, and spectroscopic probes have consistently found no sign of static order, but rather a continuum of excitations characteristic of a liquid-like state [29, 41-42].

Our dielectric and magnetic investigations place BCSO squarely in the landscape of quantum criticality. We document that BCSO indeed realizes ferroelectric quantum criticality through the $T^2$ scaling of its inverse dielectric susceptibility. Furthermore, we predict that such a ferroelectric quantum criticality originates from strong spin-correlations in the quantum critical regime, bridging between spin and lattice degrees of freedom near the quantum critical point.

The compound has been synthesised using solid-state-reaction method as described in the earlier report [42]. The high-purity precursors, $BaCO_3$, CuO, and $Sb_2O_5$, have been mixed in a stoichiometric ratio, ground, pelletized, and then calcined in air at 1070 °C with intermediate grinding and sintering. The phase purity of the sample has been characterised by performing a detailed Rietveld refinement of the X-ray diffraction (XRD) pattern of the sample, which was measured in a PanAnalytical X-ray diffractometer with Cu-$k_\alpha$ (Fig. 1). The sample forms in desired phase (space group: $P6_3/mmc$) with lattice parameters, a = b = 5.805099 Å, c = 14.321743 Å. The magnetic measurement has been performed employing a Superconducting Quantum Interference Device (SQUID) magnetometer. The dielectric constant was carried out at several frequencies (1-100 kHz) as a function of temperature down to 1.8 K using an LCR Meter (Agilent E4980A), which is integrated with commercial Physical Properties Measurement Systems (PPMS) for smooth variation of temperature.

Fig. 2 shows the inverse dielectric susceptibility $1/\varepsilon'$ plotted versus $T^2$. Over a wide temperature range, the data fall on a straight line, i.e., $1/\varepsilon' \propto T^2$, the canonical scaling expected for a three-dimensional displacive ferroelectric quantum critical point (FE QCP). This nonclassical $T^2$ scaling of the inverse dielectric susceptibility is a direct fingerprint of quantum paraelectric behaviour in which zero-point fluctuations of the polar soft mode dominate the finite-temperature response; similar extended $T^2$ ranges have been used to identify FE quantum criticality in $SrTiO_3$ and related materials [18-20]. Two further features of the dielectric data are important for the interpretation. Firstly, the quadratic scaling persists to comparatively high temperatures ~ 50 K, indicating that the polar part is already strongly renormalized by quantum fluctuations well above any common ordering temperature. High-temperature quantum fluctuations in paraelectric (dipolar) materials are reported to be significantly higher than those in low-temperature quantum fluctuations in magnetism, as observed in many other quantum paraelectric materials [18-20]. Secondly, a subtle low-temperature upturn/departure from the simple quadratic extrapolation appears below 20 K (visible in the inset of Fig. 2), suggesting the polar response begins to be affected by another low-energy degree of freedom at low T. A similar upturn in inverse dielectric constant is reported in doped-$SrTiO_3$ [44]. Both features show that the dielectric part of BCSO is very close to an FE QCP, but that other low-energy effects start to play a role at very low temperatures. Overall, the unusual $T^2$ dependence of $1/\varepsilon'$ is exactly what theory predicts for a displacive FE QCP and agrees with recent experimental findings in similar systems.

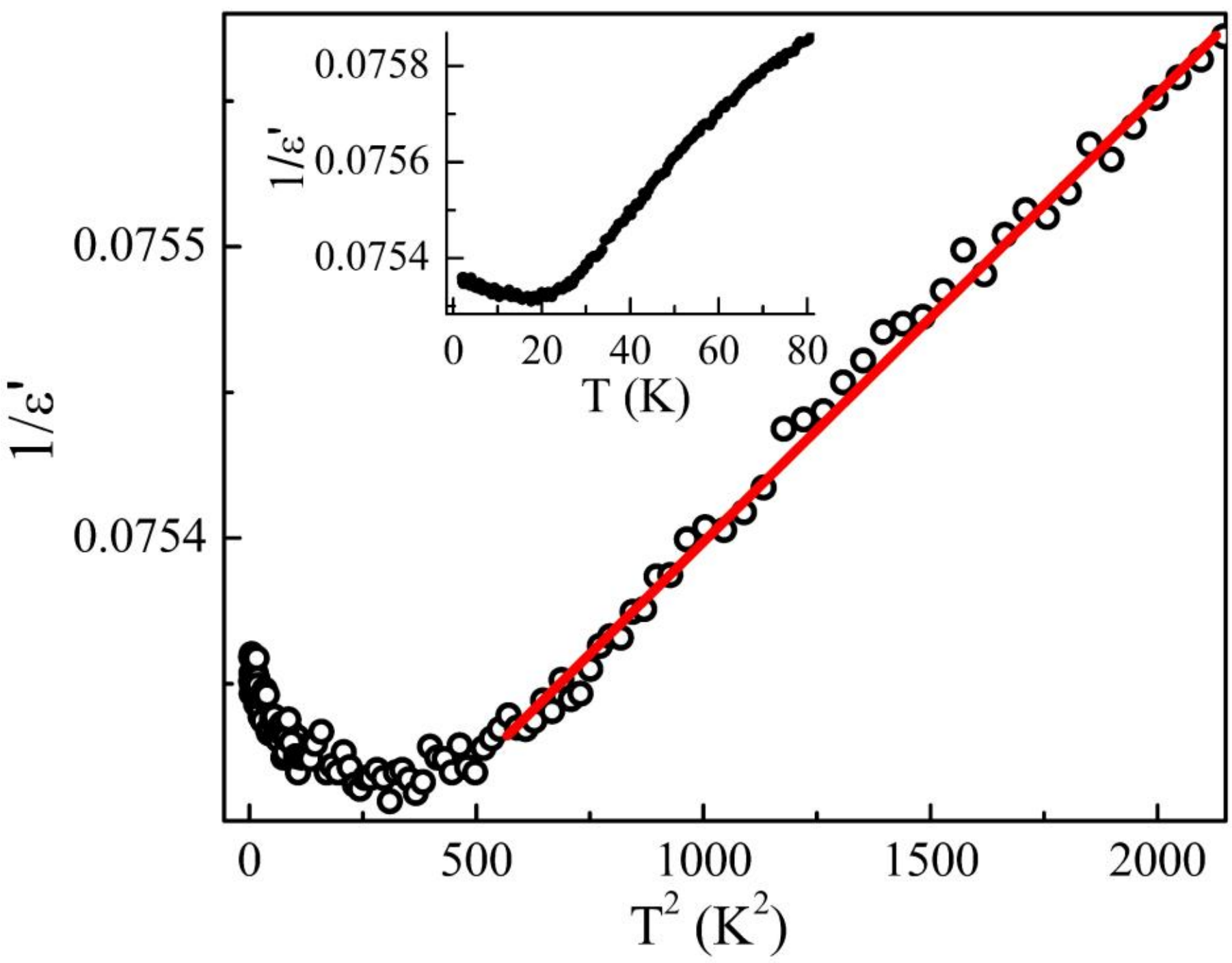

Figure 2 Inverse dielectric susceptibility vs $T^2$ plot. Inset: Inverse dielectric susceptibility vs T plot showing the deviation from the Curie-Weiss law.

The magnetic data (Fig. 3) show two crucial points, firstly, there is no sharp magnetic transition or divergence of $\chi$ down to the lowest measured temperatures 1.8 K, i.e. no conventional long-range magnetic order is detected, and secondly, the inverse susceptibility in the low-temperature window 7 K to 1.8 K follows a power law $\chi^{-1} \propto T^{3/2}$. First point places BCSO in the class of frustrated quantum magnets that avoid ordering (spin liquid or strongly dimerized/resonating states), a conclusion reached earlier by neutron, thermodynamic and local-probe studies [29, 38-42]. From the second point, the $T^{3/2}$ power law observed for $\chi^{-1}$ is the expected finite-temperature scaling of the inverse susceptibility near an antiferromagnetic (AFM) QCP in three dimensions (theoretical scaling for AFM quantum critical fluctuations gives an exponent q=3/2 under the standard Hertz-Millis type analysis for antiferromagnetic correlation) [28,44-45]. The emergence of this scaling over a finite low-temperature window is therefore strongly proposing that the spin sector is dominated by quantum critical AFM correlations rather than by classical Curie-Weiss behaviour. Similar crossovers from classical to quantum critical power laws in $\chi^{-1}$ (T) have been used previously to identify AFM quantum-critical regimes in mixed systems [28]. From the previous study, Curie-Weiss temperature $\theta_p$ = -55 K suggests that spin-correlations exist in the proximity of the FE-QCP [42].

We emphasize that the absence of long-range order (no sharp transition down to the lowest T) means that the spin system remains fluctuating; these low-energy AFM fluctuations provide a path for critical spin correlations, which can interact with other low-energy degrees of freedom, most importantly lattice or polar modes. Earlier local probes like μSR, NMR and thermodynamic work on BCSO reported singlet formation around 50 K and a continuum of low-energy magnetic fluctuations at lower T; those literatures characterize the system as either a quantum-spin-liquid or a random-singlet/dimerized state with strong low-energy dynamics, which also gives us a magnetic background compatible with the AFM-QC scaling we observe at low T [29, 33].

The coincidence of a similar quantum-critical range in both magnetic and dielectric suggests a spin-correlation-driven route to ferroelectric quantum criticality in BCSO. As the physical picture is, firstly, the dielectric response is intrinsically close to a displacive FE QCP (observed $1/\varepsilon' \propto T^2$). Secondly, this system is highly frustrated and does not order; however, it supports strong AFM quantum correlations and low-energy fluctuations (as we observed $\chi^{-1} \propto T^{3/2}$ in the low T region). Thirdly, the spin fluctuations couple to lattice/orbital degrees of freedom (via spin-phonon, magnetostriction, and JT mediated mechanisms) and can therefore renormalize the polar soft mode and favour the dielectric sector closer to a QCP or modify the low T crossover of dielectric scaling. In the previous studies, there is direct experimental evidence for strong spin-orbital-lattice entanglement in BCSO [29]. X-ray diffuse and inelastic scattering show dynamical JT distortions and short-range orbital correlations whose temperature evolution is clearly affected by the magnetic energy scale (singlet formation around 50 K), giving a direct signature that orbital or lattice degrees of freedom and spin correlations are entangled in this material [33, 39-42].

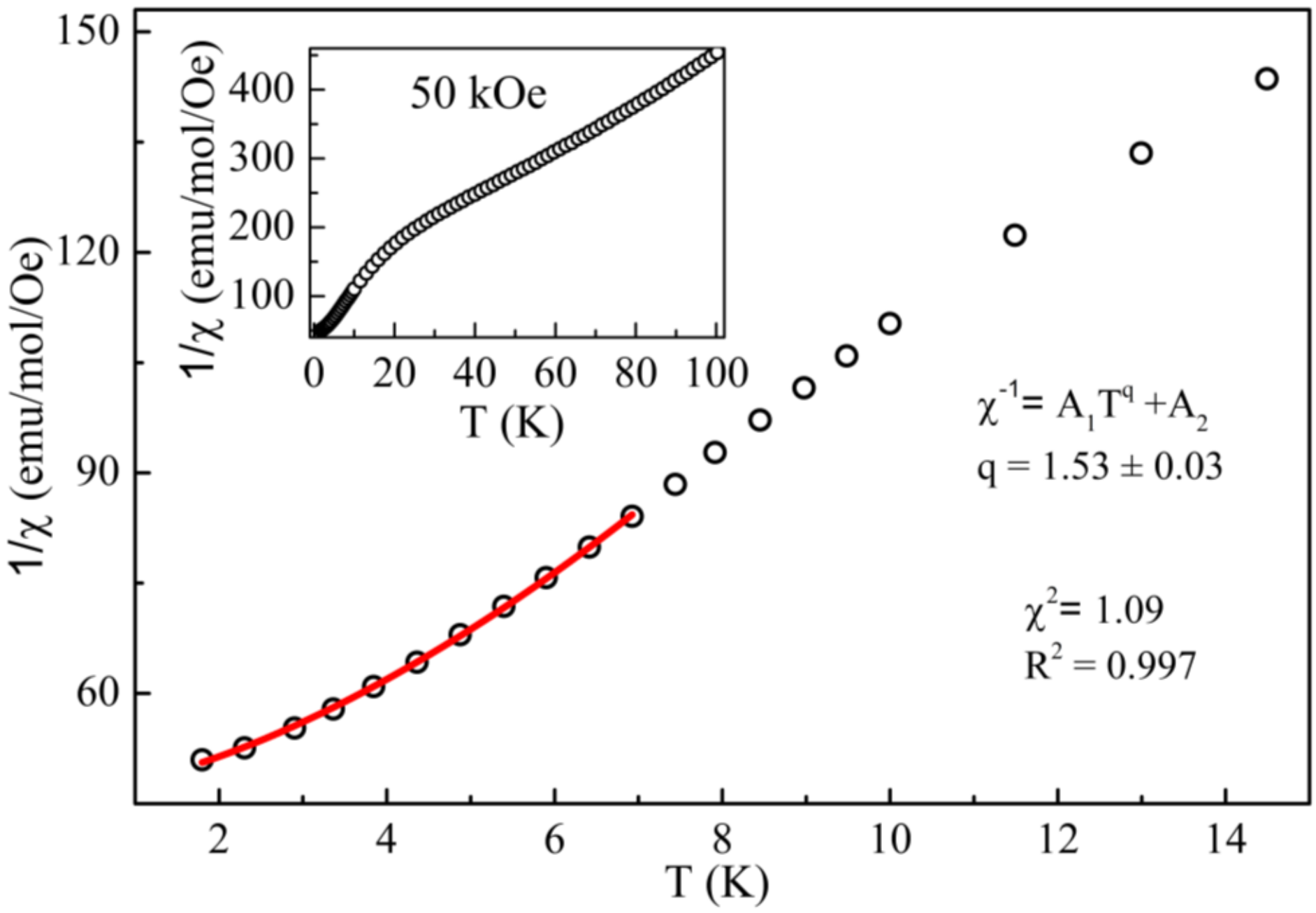

Figure 3 Inverse magnetic susceptibility vs T plot with power law fit. Inset: Inverse magnetic susceptibility vs T plot with 50 kOe field.

The observed dielectric response in our study may come from fluctuating local polar regions that arise from the strong coupling between spin, orbital, and lattice degrees of freedom in this system [46]. The $Cu^{2+}$ ($3d^9$) ions are Jahn-Teller active, and the $CuO_6$ octahedra can undergo dynamic Jahn-Teller distortions. These distortions can slightly shift the Cu ions away from the centre of the octahedra, creating small local electric dipoles that fluctuate with time instead of forming a long-range ferroelectric order. In addition, the crystal structure contains Cu-Sb dumbbells due to mutual orbital overlap [29, 46], which already have a natural charge imbalance. Any disorder or fluctuation in the orientation of these dumbbells can locally break inversion symmetry and produce local polar moments. These lattice effects can also interact with the frustrated magnetic background of the system, where short-range spin correlations may cause small displacement of the surrounding oxygen ions through spin-lattice coupling, adding another possible source of local polarization. Furthermore, the previously known spin-orbital liquid state suggests strong orbital fluctuations, which can change the local charge distribution and bonding environment, leading to additional fluctuating dipole moments [29, 40, 46]. Here, these local dipoles do not form a stable long-range ferroelectric state, their quantum fluctuations can still give rise to the dielectric quantum-critical behavior observed in our measurements. In $Ba_3CuSb_2O_9$, the $CuO_6$ and $SbO_6$ octahedra are connected through a Cu-O-Sb linkage, where the oxygen atoms mediate the interaction between the two cations. Because $Cu^{2+}$ ($3d^9$) is Jahn-Teller active while $Sb^{5+}$ ($4d^{10}$) forms a comparatively rigid octahedron, the $CuO_6$ unit tends to distort relative to the $SbO_6$ unit. This structural asymmetry

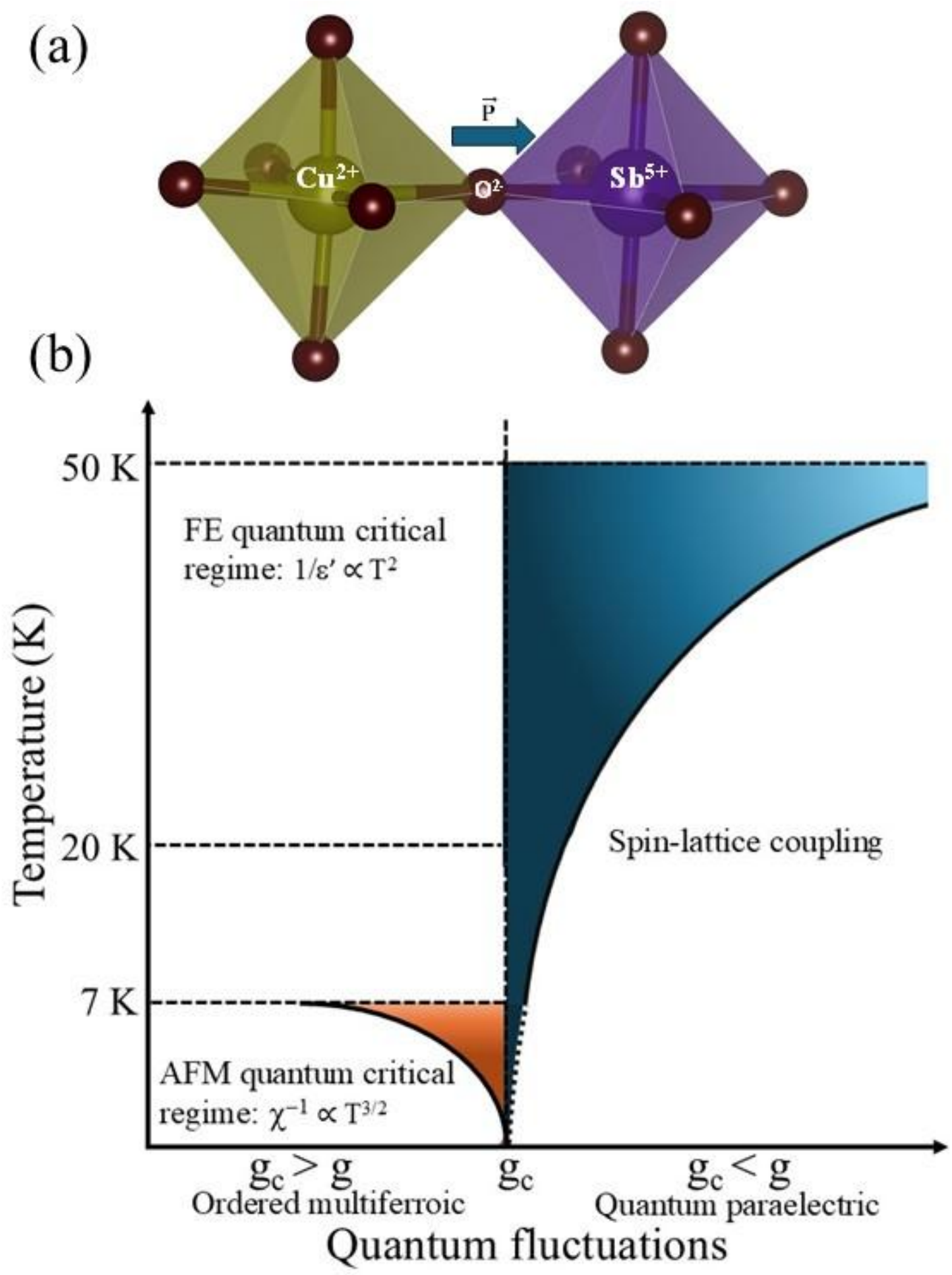

Figure 4 (a) shows the charge imbalance between the $Cu^{2+}$ and $Sb^{5+}$ ions and the local polarization P, and (b) shows the quantum critical phase diagram for BCSO.

shifts the centre of charge along the Cu-O-Sb direction, producing a local electric polarization pointing toward the Sb site (see Fig. 4a). Since the magnetic interaction between neighbouring Cu spins occurs through Cu-O-Cu superexchange paths, any change in the lattice geometry directly affects the strength of the spin interaction. In the hexagonal phase, these Jahn-Teller distortions remain dynamic rather than static, which keeps the orbitals and lattice fluctuating. As a result, the spin, orbital, and lattice degrees of freedom become strongly coupled in this system. These observations depict how magnetic fluctuations can efficiently couple to polar modes and influence ferroelectric criticality [29].

More specifically, the data are consistent with the following scenario. Above 20 K, dielectric quantum fluctuations dominate the polar response, giving the $T^2$ scaling. Below 20 K, low-energy AFM spin correlations (which become quantum critical as T is reduced toward zero) effectively renormalize the soft polar mode through spin-phonon/orbital responses; this coupling causes the small deviations from pure $T^2$ scaling visible in Fig. 2 and produces the observed AFM quantum critical signature in the magnetic response. The general idea that a critical spin correlation can modify dielectric quantum critical scaling and vice versa, that a near FE lattice can influence spin dynamics, is closely the phenomenology discussed for multiferroic quantum critical systems.

In Fig. 4b, we propose a phase diagram for spin-correlation driven ferroelectric quantum criticality in BCSO. Following the conventional quantum critical phase diagram framework, temperature is plotted on the y-axis, while the x-axis represents the strength of quantum fluctuations in a phenomenological way, replacing the usual dynamic control parameters. The diagram suggests that antiferromagnetic quantum critical fluctuations dominate below about 7 K. In the intermediate region between 7 K and 20 K, strong spin-lattice coupling develops, which subsequently drives the regime of ferroelectric quantum criticality observed between 20 K and 50 K.

Our results present a coherent picture of $Ba_3CuSb_2O_9$ in which ferroelectric quantum criticality and antiferromagnetic quantum spin correlations coexist within a strongly entangled spin-orbital-lattice framework. Earlier studies have shown the absence of magnetic order, the emergence of singlet states, and persistent low-energy fluctuations, while others have identified dynamic Jahn-Teller distortions tightly linked to magnetic behaviour. Our dielectric measurements reveal a $T^2$ dependence in dielectric constant, demonstrating the ferroelectric quantum criticality. Whereas the magnetic susceptibility follows a $T^{3/2}$ scaling typical of antiferromagnetic quantum criticality. Also, the presence of the magnetic Curie-Weiss temperature $\theta_p$ = -55 K [42], in a similar range of the FE-QCP, is a strong support for the coupling between magnetic and dielectric quantum critical response. These parallel observations suggest that neither the spin nor the lattice sector can be viewed in isolation. Instead, the collective evidence points toward a scenario where mutually coupled spin and orbital dynamics play a significant role in placing the system toward spin correlation-driven ferroelectric quantum criticality.

Till now, after theoretical prediction of multiferroic quantum criticality [14], it has been experimentally reported for the compound (Eu, Ba, Sr)$TiO_3$ [28], which is analogous to multiferroic-I, where the system is already non-polar at room temperature and further at low temperature, magnetic and ferroelectric quantum criticality coexist. Whereas this pioneering work predicts spin-correlation driven ferroelectric quantum criticality - analogous to multiferroic-II (spin-driven ferroelectricity on which spatial-inversion symmetry is broken through magnetic correlation). Such a critical phenomenon has not been experimentally reported yet. However, further focused microscopic research is warranted to understand the mechanism in detail. Considering the spin-driven ferroelectricity in $Ba_3NiNb_2O_9$ [34] and spin-correlation driven ferroelectric quantum criticality in the same family $Ba_3CuSb_2O_9$, we propose that possibly one can tune the perovskite quantum spin-liquid family through external parameters (chemical doping or applied pressure) to tune from conventional multiferroicity toward a multiferroic quantum critical state and vice-versa. In the long term, understanding and controlling such quantum critical multiferroic behavior may be useful for designing new functional materials for low-power memory devices, highly sensitive magnetic or electric sensors, and future quantum technologies.

## Acknowledgement

T.B. greatly acknowledges the Science and Engineering Research Board (SERB), now, Anusandhan National Research Foundation (ANRF) (Project No.: SRG/2022/000044), and UGC-DAE Consortium for Scientific Research (CSR) (Project No. CRS/2021- 22/03/544), Government of India, for research funding. SG thanks SERB (now, ANRF) for the fellowship. GR and EK thank RGIPT for the Institute fellowship. M.K. thanks the University Grant Commission (UGC), India, for the research fellowship. TB thanks the Central Instrumentation Facilities (CIF), RGIPT. The authors thank Prof. E. V. Sampathkumaran, Tata Institute of Fundamental Research (TIFR), Mumbai, India, for giving access to some of the experimental facilities. The authors thank Prof. S. Majumder, IACS, for fruitful discussions.